\documentclass[a4paper,12pt,final]{article}
\usepackage{t1enc}
\usepackage[english]{babel}
\usepackage[utf8]{inputenc}
\usepackage{floatflt}
\usepackage{graphicx}
\usepackage{psfrag}
\usepackage{bbm}
\usepackage{amsmath}
\usepackage{amssymb}
\usepackage{showkeys}
\usepackage{hyperref}
\usepackage{ifthen}
\usepackage{subfigure}
\usepackage{epstopdf}

\title{Vortices with scalar condensates in two-component Ginzburg-Landau systems}
\author{Péter Forgács\textsuperscript{1,2}, Árpád Lukács\textsuperscript{1}\\
{\small {}\textsuperscript{1} Wigner RCP RMI, H1525 Budapest, POB 49}\\
{\small {}\textsuperscript{2} LMPT CNRS UMR7350, Universit\'e de Tours, Parc de Grandmont, 37200 Tours, France}
}

\def\d{\mathrm{d}}
\def\e{\mathrm{e}}

\def\kihagy#1{}

\newcommand{\arxiv}[2][]{
  \ifthenelse{\equal{#1}{}}{
    \href{http://arxiv.org/abs/#2}{\texttt{arXiv:#2}}
  }{
    \href{http://arxiv.org/abs/#2}{\texttt{arXiv:#2 [#1]}}
  }
}

\newcommand{\be}{\begin{equation}}
\newcommand{\ee}{\end{equation}}

\begin{document}
\maketitle

\begin{abstract}
In a class of two-component Ginzburg-Landau models (TCGL) with a U(1)$\times$U(1) symmetric potential, vortices with a condensate at their core may have significantly lower energies than the Abrikosov-Nielsen-Olesen (ANO) ones. On the example of liquid metallic hydrogen (LMH) above the critical temperature for protons we show that the ANO vortices become unstable against core-condensation, while  condensate-core (CC) vortices are stable. For LMH the ratio of the masses of the two types of condensates, $M=m_2/m_1$ is large, and then as a consequence the
energy per flux quantum of the vortices, $E_n/n$ becomes a non-monotonous function of the number of flux quanta, $n$. This leads to yet another manifestation of neither type 1 nor type 2, (type 1.5) superconductivity: superconducting and normal domains coexist while various ``giant'' vortices form. We note that LMH provides a particularly clean example of type 1.5 state as the interband coupling between electronic and protonic Cooper-pairs is forbidden.
\end{abstract}

Multi-component Ginzburg-Landau models have received much attention recently \cite{ZD, BabaevF, BS, BCSt15, BJS, Zou} as they provide reasonably good phenomenological description of various multi-band superconductors, such as the celebrated magnesium diboride (MgB${}_2$). Importantly it has been shown in Refs.\ \cite{ZD,BS1, BS2} that two-component Ginzburg-Landau models (TCGL) do provide an effective description of superconductors  with two order parameters starting with the microscopic theory. The TCGL description is also very useful for more exotic matter, when pairings of different particles occur, a prominent example being liquid metallic hydrogen (LMH) \cite{Ashcroft68, Ashcroft91, Ashcroft98,
Ashcroft99,Ashcroft2000, Ashcroft2005,SBP}. LMH is expected to be created under sufficiently high pressure, when electron, resp.\ proton pairing would take place below two distinct critical temperatures. For the state of the experimental search for LMH, see Refs.\ \cite{ET,hardpressed}, and Ref.\ \cite{SSBS,simul} for computer simulation results. Moreover the multicomponent GL theory is also of relevance for the description of neutron star interiors when pairings of different particle species are also expected to occur \cite{Jones}.
As the magnetic properties of superconductors are determined to a large extent by its vortices and the interactions between them \cite{Abrikosov, FS, SBP, Pismen}, they are of utmost importance.

TCGL models generically possess a U(1)$\times$U(1) symmetric potential for the two scalar fields (order parameters), with two different correlation lengths. In such theories the physics of vortices is richer than that of the Abrikosov-Nielsen-Olesen vortices in the GL model \cite{Abrikosov,NO}. For example when the value of the  magnetic penetration depth falls between the two correlation lengths, inter-vortex forces are in general attractive for small vortex-separations while being repulsive for larger ones, see Refs.\ \cite{BS, CGB, GB} where quite remarkable multi-vortex states have been observed. The non-monotonous nature of the inter-vortex forces is associated with what has been dubbed as type 1.5 superconductivity \cite{Moshchalkov, BS}. Moreover vortices in TCGL can carry fractional magnetic flux \cite{BabaevF}.
In most superconductors the two order parameters are coupled due to interband interactions, and there is a single critical temperature. Then below the critical temperature both condensates are superconducting, i.e.\ both order parameters have a non-zero vacuum expectation value (VEV). Previous studies have mostly concentrated on this 2VEV case.

In the high-energy physics context, flux-tube or string solutions of the relativistic version of TCGL coupled to another gauge field -- a U(1)$\times$U(1) Higgs model -- interpreted as superconducting cosmic string has been shown to be of great interest \cite{Witten}. The $SU(2)$ symmetric version of the (relativistic) TCGL theory, known as the SU(2)-semilocal model in the literature \cite{vac-ach} corresponds to the
limit of the bosonic sector of the standard model, when the Weinberg mixing-angle, $\theta_W\to\pi/2$. 
In the semilocal model because of the global SU(2) symmetry one can always achieve that only one of the order parameters
takes on a VEV. Semilocal vortices exhibit a number of remarkable features, e.g.\ there exists a 1-parameter family of stable vortex solution at the border between type 1 and type 2 cases (Bogomolny point) corresponding to a deformation of the embedded ANO vortex, with a condensate at its core \cite{vac-ach, hin1, hin2, semilocal}. Condensate core (CC) vortices
have been considered in the semilocal model away from the Bogomolny point in Refs.\ \cite{FRV1, FRV2}, and for an overview for the case of a general U(1)$\times$U(1) symmetric potential we refer to Ref.\ \cite{Erice}.
More recently CC vortices have been investigated  in Ref.\ \cite{GB} with a special, $U(1)\times U(1)\times \mathbb{Z}_2$ symmetric potential, when both vortices and domain walls coexist, and even a hybrid vortex-domain wall type of excitation may appear.

The aim of this paper, is to present the some interesting physical phenomena due to CC vortices in TCGL models with
U(1)$\times$U(1) symmetric potentials. We shall concentrate on the case when only a single scalar field has a non-zero value in the minimum of the potential, i.e. when there is a single VEV.
In condensed matter systems the scalar potential can be of this 1VEV-type, when there is no Josephson coupling between the two order parameters, a prominent example being LMH. In LMH pairings between electrons and between protons are possible, when the temperature of the system is between the electronic and the protonic critical temperature, since the two type of Cooper pairs have a different charge forbidding interband coupling.
ANO vortices can be embedded in TCGL models when only one of the fields takes on a VEV, by setting the VEV-less field to zero. We demonstrate that the CC vortices (with a condensate at their core) are stable, as opposed to the embedded ANO ones which are unstable. In CC vortices the field with 0 VEV takes on a non-zero value \emph{in the vortex core}, lowering the energy of the vortex, and stabilizing it. Depending on the value of the parameters, the energy of CC vortices can be significantly lower than that of the corresponding ANO ones.
 We find that there are significant physical implications of the second order parameter in the case when the ratio of the masses of the two types of condensates, $M=m_2/m_1$, is large. For LMH this ratio can be considered to be large, 
 $M=m_{\rm proton}/m_{\rm electron}\sim1836$. When $M$ is large the energy per flux quantum of the vortices, $E_n/n$ becomes a non-monotonous function of the number of flux quanta, $n$. What happens is that for a given value of $M$ the energy/flux, $E_n/n$ decreases, so ``giant'' vortices (with a large flux) with a considerable amount of flux quanta may form, however, there is always maximal value of $n$ for which the binding energy becomes zero. This is a rather neat manifestation of type 1.5 behaviour. The magnetic properties of such a superconductor do not fall into the classification of either type 1 or type 2: although vortices tend to stick together to form ``giant'' vortices, various normal and superconducting domains coexist.

The simple two component Abelian model can be applied in the context of Higgs-portals \cite{SilveiraZee,PW}. In that case the scalar field with 0 VEV is neutral under the U(1) and it couples only through the symmetry breaking field to ordinary matter.
In this context we have a simplified model of CC cosmic strings when dark matter is bound to the cosmic string core. The model considered is a simplified version of the scalar phantom (or Higgs portal) model of dark matter \cite{SilveiraZee,PW}.

\section{The two component Ginzburg-Landau theory}
In its nondimensionalised form, the energy density of a two-component Ginzburg-Landau (TCGL) model is
\begin{equation}\label{eq:erg}
E = \int \d^3 x \mathcal{E}\,,\quad \mathcal{E} = \sum_{a=1}^2 |{\bf D}\phi_a|^2 + \frac{|{\bf B}|^2}{2} + V(\phi_a,\phi_a^*)\,,
\end{equation}
where ${\bf D}\Phi_a=(\nabla-e_a {\bf A})\phi_a$, with $e_a$ the charge of the condensate $a$, $a=1,2$ and ${\bf B}=\nabla\times {\bf A}$.
The most general fourth-order potential with $U(1)\times U(1)$ symmetry is \cite{Witten}
\begin{equation}
  \label{eq:pot}
  V = \frac{\beta_1}{2}\left(|\phi_1|^2-1\right)^2 + \frac{\beta_2}{2}\phi_2^4 - \alpha |\phi_2|^2 + \beta'|\phi_1|^2|\phi_2|^2\,.
\end{equation}
The parameterisation of the potential was chosen in a way most suitable for the
study of the 1VEV state, with $\phi_1$ having a non-zero VEV.

For the stability of the potential,
\begin{equation}\label{eq:stabcond}
\beta_{1,2}>0, \quad \beta'>-\sqrt{\beta_1 \beta_2} \,.
\end{equation}
must hold. Depending on the parameters of the potential, the
minimum is either at $|\phi_1|=\eta_1$, $\phi_2=\eta_2$, $\eta_{1,2}\ne 0$ (2VEV case),  at $|\phi_1|=1$ $\phi_2=0$ or at $\phi_1=0$, $\phi_2=\eta_2$ (1VEV cases).
In the present paper, we consider vortices in the upper component 1VEV case, for which
\begin{equation}
 \label{eq:1VEVcond1}
  \beta_1 \beta_2 > {\beta'}^2\,,\text{and}\ \beta'>\alpha,
\end{equation}
or
\begin{equation}
 \label{eq:1VEVcond2}
  \beta_1 \beta_2 < {\beta'}^2\,,\text{and}\ \sqrt{\beta_1\beta_2}>\alpha
\end{equation}
must hold.

Previous studies have focused on the 2VEV case, either in a condensed matter \cite{BabaevF,BS,BJS, BCSt15, CBS, BSAdet, BS1, BS2}
or a high energy physics \cite{Witten, Peter, Peter2}. The $SU(2)$ symmetric model, called semilocal has been considered in Refs.\ \cite{vac-ach, hin1, hin2, semilocal, FRV1, FRV2, twistedinstab1, FL, twistedinstab2}.

The parameters of the potential are $\beta_1 = 4\lambda_1 m_1^2/(\hbar^2e^2\mu_0)$,
$\beta_2 = 4\lambda_2 m_2^2/(\hbar^2 e^2\mu_0)$, $\alpha = 4 \nu_2 m_1 m_2 /(\hbar^2 e^2 \mu_0 \eta_1^2)$,
$\beta'=4 \lambda' m_1 m_2/(\hbar^2 e^2 \mu_0)$, where $\lambda_{1,2}$, $\lambda'$ and $\nu_{1,2}$ are the expansion coefficients of  the Ginzburg-Landau
free energy, $\mu_0$ is the vacuum permeability, $e_i$ the charge of the condensate $i$ measured in units of $e$, a charge unit (for Cooper pairs,
conveniently, twice the elementary charge), $\eta_1=\sqrt{\nu_1/\lambda_1}$. In these units, the value of $\phi_1$ at the minimum of the potential is scaled
to 1, the penetration depth, $\lambda_L=\sqrt{m_1/(\mu_0 e_1^2e^2\eta_1)}$ is scaled to $\sqrt{2}$. With the mass ratio of the two kinds of Cooper pairs,
$M=m_2/m_1$, the parameters scale as $\beta_2 = M^2{\tilde{\beta}}_2$, $\beta'=M{\tilde{\beta}}'$, $\alpha=M\tilde{\alpha}$, where the tilde quantities are
expected to be of the same order of magnitude. For LMH, $e_1=-e_2=1$, and the $M\gg 1$ limit is of interest.

\section{Properties of vortices with condensate in their cores}\label{sec:vort}
We seek vortex solutions with the help of the rotationally symmetric Ansatz,
\begin{equation}
  \label{eq:Ans}
\begin{aligned}
  \phi_1(r,\vartheta,z)\, &= f_1(r) \e^{i n\vartheta}\,, \\
  \phi_2(r,\vartheta,z)\,  &= f_2(r)\e^{i m\vartheta}\,, \\
\end{aligned} \quad
A_\vartheta(r,\vartheta,z) = n a(r)\,,
\end{equation}
with $A_0=A_3=0$, $A_r=0$, where $r,\vartheta,z$ are cylindrical coordinates, and the vortex line is aligned with the $z$ axis.
In the 1VEV case, solutions of the standard GL theory can be trivially embedded in the two-component theory with $\phi_1=\phi$, $\phi_2=0$.
The embedded ANO-vortex, depending on the parameters of the potential, may be
stable or unstable. The case, where the potential was $SU(2)$ symmetric, has been considered in Refs.\ \cite{vac-ach,hin1, hin2,semilocal}.
In the $SU(2)$ symmetric case, the embedded Abrikosov vortex is unstable if the GL parameter is $\beta >1$.  We have similarly
linearised the field equations in $\phi_2$. We have found, that the ANO vortex is unstable for a wide range of the parameters,
e.g., close to the $SU(2)$ symmetric case, or for $\beta_1 >1$ and $\beta_2$ large, with $\beta'$ and $\alpha$ chosen appropriately in between.

Vortex solutions were computed by numerically solving the radial equations obtained by plugging the Ansatz (\ref{eq:Ans}) in the energy density
(\ref{eq:erg}), using code from Ref.\ \cite{NR}. We have found, that if the ANO vortices are unstable and if $\beta' > \alpha$,
there exists a solution for which  $\phi_2$ does not vanish in the vortex core. These vortices also exist in the case, when
$\beta'=\alpha$ (the boundary between the 1VEV and 2VEV cases), if $\beta_1\beta_2\ne{\beta'}^2$, however, not exponentially localised in this case. Continued into
the range $\alpha > \beta'$ (2VEV), these vortices become the fractional flux vortices of Refs.\ \cite{BabaevF, BS}.
If $\beta_1\beta_2 = \alpha^2$ (the boundary between upper and lower component 1VEV), solutions with the upper and the lower component having a non-zero VEV exist.
For more details, see also Ref.\ \cite{Erice}. For the special case of $U(1)\times U(1)\times\mathbb{Z}_2$ symmetry, see Ref.\ \cite{GB}. The domain structure observed there
is a consequence of the high degree of symmetry of their potential.

\emph{The CC vortices have significantly lower energy than ANO ones} (see Table \ref{tab:erg}), 
and this difference grows rapidly with the mass ratio $M$. They are also \emph{stable at the linear level}: we have repeated the stability analysis of Ref.\ \cite{FL}
(with the methods of Ref.\ \cite{Goodband}), and found that for CC vortices, the perturbation equations have no negative eigenvalues.

An analytical approximation is possible for $n\to\infty$ and $e_2=0$. ANO vortices approach a false vacuum bag with radius $R_{A}\sim \sqrt{2 n}\beta^{-1/4}$
and energy $E_{A} \sim 2\pi n \sqrt{\beta}$ \cite{Bolognesi1, Bolognesi2}. For CC vortices, we obtained radius $R_C \sim \sqrt{2 n}(\beta_1-\alpha^2/\beta_2)^{-1/4}$
and energy $E_C \sim 2\pi n \sqrt{\beta_1 -\alpha^2/\beta_2}$. To the $e_2=1$ case, we shall return in Section \ref{sec:LMH}.

\begin{table}
 \begin{center}
 \begin{tabular}{|c||c|c|c||c|}
    \hline
    $n$            & (a)      & (b)        & (c)        & ANO\\
    \hline
    1              & 1.152   & 1.008       & 0.78       & 1.157 \\
    2              & 1.121   & 0.913       & 0.75       & 1.210 \\
    3              & 1.107   & 0.882       & 0.72       & 1.239 \\
    \hline
 \end{tabular}
  \end{center}
  \caption{Energy per unit flux, $E_n/(2\pi n)$ of CC vortices for (a) $\beta_{1,2}=\alpha=2$, $\beta'=2.1$, (b) $\beta_1=2$, $\beta_2=8$, $\beta'=4.2$, $\alpha=4$
  and (c) $\beta_1=2$, $\beta_2=3872$, $\beta'=87.4$, $\alpha=83$
  compared to ANO $\beta=2$.}
  \label{tab:erg}
\end{table}

\section{Vortices in superconducting LMH}\label{sec:LMH}
\begin{figure}[h!]
 \noindent\hfil\includegraphics[angle=-90,scale=.5]{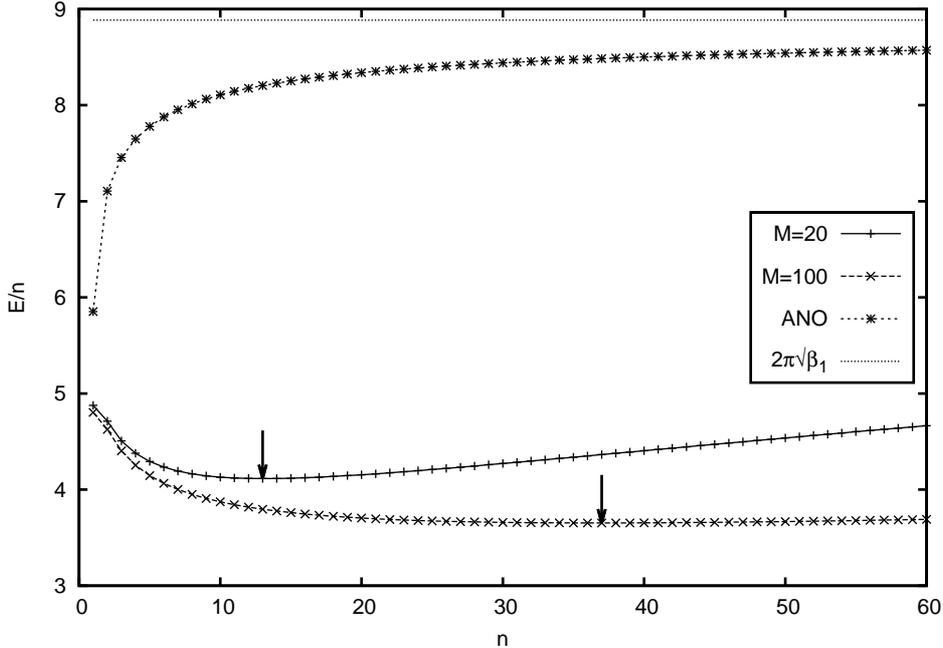}
 \caption{Energy of vortices per unit flux, $\beta_1=2$, $\tilde{\beta}_2=9.68$, ${\tilde{\beta}}'=4.37$, $\tilde{\alpha}=4.15$ and different values of $M$
 compared to Abrikosov (ANO) vortex energies. The arrows mark the minima of the energy per flux.}
 \label{fig:En}
\end{figure}

For LMH, the $M\gg 1$ limit is relevant. In Fig.\ \ref{fig:En}, we have plotted the energy of vortices vs.\ the number of their flux quanta, for $M=20$ and $M=100$. Note first, that there is quite a large difference between the energies of CC and ANO vortices. The second important feature of the results is that $E_n/n$ has a minimum, which is shifted to higher number of flux quanta, $n$, with higher mass ratio $M$. As a consequence, single flux vortices attract each other (similarly to type I SCs), however, for very large $n$,
there is a repulsive interaction (similarly to the case of type II SCs). In between, there is a vortex which has the strongest binding.
The magnetic behaviour of the material is thus different from both type I and type II superconductors: in an external magnetic field, instead of an Abrikosov lattice of unit flux vortices, vortices with quite large (several hundred) flux quanta form, many close to the minimum of $E_n/n$, but some significantly larger (possibly a few thousand flux
quanta). However, the latter do not unite to form normal domains. For $M=1000$, we have used an approximate configuration to obtain the minimum, at $n\approx 350$, and the last bound vortex at $n\approx 6600$. For lower values of M, we have numerical data, e.g., for $\beta_1=2$, ${\tilde{\beta}}_2=9.68$, ${\tilde{\beta}}'=4.37$ and
${\tilde{\alpha}}=4.15$ for $M=20$ the minimum is at $n=13$, $E_1/(2\pi)=0.7762$, and $E_{13}/(26\pi)=0.6549$. The maximum bound vortex, i.e.,
where $E_n/n$ reaches $E_1$ is $n=78$. For $M=100$, $E_1/(2\pi)=0.7646$ and the minimum is at $n=37$, and $E_{37}/(74\pi)=0.5814$, and the last bound
vortex is around $n=380$.

The behaviour when $n$ and $M$ are both large can be also understood with the help of a configuration. Let $f_1$ be zero in the core, and $f_2$
approximately $\sqrt{\alpha/\beta_2}$ there. Between $(1-\delta )R$ and $R$, $f_1$ is linear, and reaches 1 at $R$, and $f_2$ also linear, and reaches 0.
For $r<R$, $a=(r/R)^2$, and for $r>R$, $f_1=1$, $f_2=0$ and $a=1$. An approximate minimalisation yields
$\delta \approx (5/2)^{1/4}((\alpha/\beta_2+M)/n^2/(M-3e_2^2\alpha/\beta_2))^{1/4}$, $R\approx\sqrt{2n}(\beta_1-\alpha^2/\beta_2)^{1/4}$
and
\begin{equation}\label{eq:nMappr}
 E \approx \frac{e_2^2\pi{\tilde{\alpha}}}{2\tilde{\beta}_2 M}n^2 + 2\pi\sqrt{\beta_1-\frac{{\tilde{\alpha}}^2}{\tilde{\beta}_2}}n
 +\frac{8\pi\sqrt{n}}{3}\left(\frac{2}{5}\right)^{1/4}\,.
\end{equation}
For $e_2=1$, at a given $n$, this approximation yields larger energies than the Bolognesi bag \cite{Bolognesi1, Bolognesi2}, i.e., for $n\to\infty$, this variational
Ansatz breaks down, and the vortex energy approaches the same asymptotic form as the ANO energy, $2\pi n \sqrt{\beta_1}$. For moderate values of $n$,
formula (\ref{eq:nMappr}) describes the energy of numerical solutions qualitatively well.


Let us consider some special cases here: the boundary between the parameter ranges corresponding to 1VEV and 2VEV is at $\alpha=\beta'$, while
there is a boundary between upper and lower component 1VEV at at $\alpha=\sqrt{\beta_1\beta_2}$, if $\beta' > \alpha$ remains valid.

Close to the former one, $\alpha=\beta'$, the second condensate in the vortex core expands, and its radial falloff, $\sim \exp(-\sqrt{\beta'-\alpha}r)$
becomes slower. When the characteristic length scale of the falloff reaches the sample size, these vortices become indistinguishable from
the fractional flux vortices of Ref.\ \cite{BabaevF, BS} in the 2VEV case. For $\alpha > \beta'$, vortices with winding in the lower component also appear.

Close to $\alpha=\sqrt{\beta_1 \beta_2}$, the potential energy in the core becomes small, the vortices become large, and their flux is localised closer to
the outer end of their cores. At the same time, the minimum of $E_n/n$ is shifted to larger values of $n$, and at $\alpha=\sqrt{\beta_1 \beta_2}$,
$E_n \propto n$ for large $n$. Here, ANO vortices in the lower component become also allowed.
In this case, it is possible to exchange the role of the 2 components, with the rescaling $\phi_a\to \eta_2\phi_a$, $x\to x/\eta_2$, $A\to \eta_2 A$, where $\eta_2^2=\alpha/\beta_2$.
In this way, we get the same expression (\ref{eq:erg}, \ref{eq:pot}) for the energy of the vortices with an overall multiplier $\alpha/\beta_2$.
With the same configuration as above, the estimated energy of these vortices is $E=2\pi( 4\alpha/\beta_2 + \alpha/\sqrt{3\beta_2})$, which
is $M^0$ asymptotically. However, using the large-$\beta$ asymptotics of Abrikosov vortex energy \cite{Pismen}, we get $E\sim 2\pi\frac{\alpha}{\beta_2}\log\sqrt{\beta_2}$,
i.e., $\sim (\log M)/M$, telling us that at the transition, it is energetically favourable for the vortices to break up into $n=1$ lower component
Abrikosov vortices. Linearising the equations in the other component shows, that these vortices are then stable against the formation of a condensate
in their core. This can be seen as follows: the large-$\beta$ asymptotic form of the vortex profile is a small core
with size proportional to $1/\sqrt{\beta_2}\propto 1/M$. The linearised equation is of the form of an eigenvalue equation,
and we have verified numerically, that it has no bound modes, and therefore if $\alpha > \sqrt{\beta_1 \beta_2}$, there is ordinary superconductivity.


\section{Implications for electroweak strings}
In the semilocal model, $\beta_1=\beta_2=\beta'=\alpha \equiv \beta$, embedded ANO strings are unstable for $\beta >1$ \cite{JPV1, JPV2, Perkins, vac-ach, hin1, hin2,
GHelectroweak, semilocal, VS, kibble}. The effect of an additional scalar field on the stability of vortices has been considered in Ref.\ \cite{VachaspatiWatkins} for $\beta_2=0$.
The resulting model is the one considered here with the field $\phi_1$ promoted to a two-component field $\Phi_1$.
Extending the study to $\beta_2 >0$, it is found that the range of stability of the vortices is considerably extended. A candidate for the role of the second scalar field
would be the scalar phantom of Refs.\ \cite{SilveiraZee, PW}, a dark matter candidate, however, the the parameter range where stabilisation occurs, is phenomenologically disfavoured
\cite{Zee1, Zee2, Beniwal}. With the added stabilising effects of dark scalars, and possibly other fields, such as the dilaton \cite{PerivolaropoulosPlatis}
or a dark $U(1)$ with gauge kinetic mixing \cite{HartmannArbabzadah, BrihayeHartmann}, the question of finding stable semilocal or electroweak strings remains open.
 
\section{Conclusions}
In type II SCs ($\beta>1$), inter-vortex forces are repulsive, implying $E_2 > 2E_1$. In the two-component system, depending
on the parameters, e.g., in the $M\gg 1$ limit, for a range of $n$, $E_n < n E_1$, which implies now, that there is a range of distances,
where inter-vortex forces are attractive, i.e., the addition of the second component leads to a drastically changed and richer physics,
similar to that of type 1.5 SCs.

Depending on relative magnitude of the parameters $\sqrt{\beta_1\beta_2}$,  $\alpha$, and $\beta'$, there are different adjacent states
in the systems: 1VEV and 2VEV or upper and lower component 1VEV). In the latter case, at the transition, CC vortices become large, the second
condensate pushes out the flux into a wall for large $n$ vortices, and then this wall breaks up into Abrikosov vortices in the second condensate, having significantly lower energy
in the new phase, and being stable against condensate formation in their core.

If $\sqrt{\beta_1\beta_2} > \beta'$, there is a 1VEV to 2VEV transition, and a second critical temperature, below which both condensates
are active. At the second critical temperature, the new vortices expel some of their flux, and approach the fractional flux vortices
put forward by Refs.\ \cite{BabaevF, BS}. At the critical temperature, the radial decay of the second component becomes slow,
and our vortices approach the fractional flux vortices of Refs.\ \cite{BabaevF,BS}. As the transition is approached,
vortex interaction become attractive and large flux vortices are expected to form. After the transition,
there are non-trivial multi-vortex structures for large $n$ \cite{CGB}.

\subsection*{Acknowledgements}
This work has been supported by the grants OTKA K101709. The authors would like to thank Prof.~F.A.~Schaposnik for discussions.

\def\refttl#1{{\sl ``#1''}, }%

\end{document}